\documentclass[pra,twocolumn,superscriptaddress,showpacs]{revtex4-1}
\usepackage{amsbsy}
\usepackage{amsfonts}
\usepackage{amssymb}
\usepackage{amsmath}
\usepackage{graphicx,color}
\usepackage{graphicx}

\begin{document}

%opening

\title{Equilibrium and Disorder-induced behavior in Quantum Light-Matter Systems}

\author{Eduardo Mascarenhas}
\affiliation{Departamento de F\'isica, Universidade Federal de Minas Gerais, Belo Horizonte, MG, Brazil}

\author{Libby Heaney}
\affiliation{Department of Physics, University of Oxford, Clarendon Laboratory, Oxford, OX1 3PU, United Kingdom}
\affiliation{Centre for Quantum Technologies, National University of Singapore, Singapore}

\author{M. C. O. Aguiar}
\affiliation{Departamento de F\'isica, Universidade Federal de Minas Gerais, Belo Horizonte, MG, Brazil}

\author{Marcelo Fran\c{c}a Santos}
\affiliation{Departamento de F\'isica, Universidade Federal de Minas Gerais, Belo Horizonte, MG, Brazil}

\begin{abstract}

We analyze equilibrium properties of coupled-doped cavities described by the Jaynes-Cummings-Hubbard Hamiltonian. In particular, we characterize the entanglement of the system in relation to the insulating-superfluid phase transition. We point out the existence of a crossover inside the superfluid phase of the system when the excitations change from polaritonic to purely photonic. Using an ensemble statistical approach for small systems and stochastic-mean-field theory for large systems we analyze static disorder of the characteristic parameters of the system and explore the ground state induced statistics. We report on a variety of glassy phases deriving from the hybrid statistics of the system. On-site strong disorder induces insulating behavior through two different mechanisms. For disorder in the light-matter detuning, low energy cavities dominate the statistics allowing the excitations to localize and bunch in such cavities. In the case of disorder in the light-matter coupling, sites with strong coupling between light and matter become very significant, which enhances the Mott-like insulating behavior. Inter-site (hopping) disorder induces fluidity and the dominant sites are strongly coupled to each other.

\end{abstract}
\maketitle

\section{Introduction}

Quantum phase transitions are a remarkable zero temperature phenomenon driven by quantum fluctuations~\cite{QFT}. Such transitions have been studied in many-body quantum systems where each quantum phase can be unambiguously defined. However, recent results show evidence that interesting aspects and important traces of the physics of novel quantum phase transitions may already be observed in the limit of very few interacting sites~\cite{Bose,EN, ElenGround,Small1,Small}. This is particularly clear in hybrid light-matter systems, such as coupled electromagnetic cavities doped with two level impurities, where a Mott insulating to superfluid crossover has been predicted for as few as six or seven sites~\cite{Bose}. These systems feature a composite fermion-boson excitation in each site, hence the term \textit{hybrid}, and quantities like the variance in energy for each site have been used as markers for the transition between different phases~\cite{DMRGCavs}. 
However, entanglement, an unique quantum correlation with no classical analog which has been related to fundamental features of quantum phase transitions~\cite{CriticalEnt}, may be regarded as a more adequate order parameter~\cite{OrdeEnt, ElenGround}. In this work, we study the system entanglement to show how such quantum correlations relate to the behaviors the system may present.

One possible Hamiltonian describing doped and coupled cavities is the Jaynes-Cummings-Hubbard (JCH) model~\cite{Bose, JCHM} which, in some limiting approximations, mimics the more typical and simpler Bose-Hubbard one. The similarities between both models have prompted the use of the latter as a basis for the analysis of quantum phase transitions in the former both for a large~\cite{Dario} and a very small number of sites~\cite{Dario2}. However, the analogy to this simpler model ignores the internal structure of each site what prevents one from exploring the increased complexity of the JCH system. 
The implementation of such systems has been proposed in different quantum optical setups such as planar lattices of one mode cavities each containing one quantum dot~\cite{Michal}, photonic crystal microcavities~\cite{Yamamoto}, circuit quantum electrodynamics with a finite system approach~\cite{Dario3}, and in trapped ions~\cite{SimulIon}. One of the greatest advantages of all these setups is the combination of highly controllable experimental conditions and the large effective size of each site that allows for the design of mesoscopic simulators of condensed matter systems. 

In many cases the JCH system is naturally disturbed by noise that usually takes the system out of equilibrium. However, even in equilibrium, disordered imperfections in the system preparation may induce transitions that drastically change quantum phases and their correlations. 
Disorder may manifest itself in very different and even opposite effects. The lattice imperfections, that differ from site to site, may suppress quantum coherence inducing the spatial localization of quantum states destroying the system fluidity, which leads to compressible, despite non-fluid, glassy phases~\cite{Glass}. However, disorder may induce fluidity under certain circumstances~\cite{Small,DIfluid,SMFT2}. The hybrid nature of the system also leads to interesting effects under the action of disorder, as we show in the following sections.

In this paper, we show that the entanglement between different constituents of the JCH system can be used not only to characterize the already known phase transitions, also present in the Bose-Hubbard model, but also, and more importantly, to identify new behavior involving the nature, either hybrid or bosonic, induced by the more complex JCH interaction. We address small and large quantum systems, extremes that present similarities and differences that are of great interest: while few sites are experimentally feasible in a controllable way, phase transitions are better defined in large samples. 
We also analyze the entanglement and the disorder-induced effects of the JCH hamiltonian. For the analysis of the small system we enter deeply in the statistics of the ensembles induced by disorder, since in principle a physical observer could perform spectroscopic measurements of the system structure and obtain the disordered pure states (or at least quasi-pure) pertaining to the induced ensembles. For the analysis of the large system we resort to stochastic-mean-field-theory (SMFT), which was recently developed in~\cite{SMFT,SMFT2} and allows us to study on site statistics. We show how the statistics of the system change under the various ways in which disorder may set in and also show the disorder-induced phase transitions. 

The analysis of the clean system is developed in section II with one subsection for the small limit and another for the large limit. The disordered small system is addressed in section III and the disordered large system is addressed in section V after a recollection of SMFT in section IV. Section VI concludes the paper.

\section{The Hybrid System: Jaynes-Cummings-Hubbard Hamiltonian}

\begin{figure}[h]
\includegraphics[width=8cm]{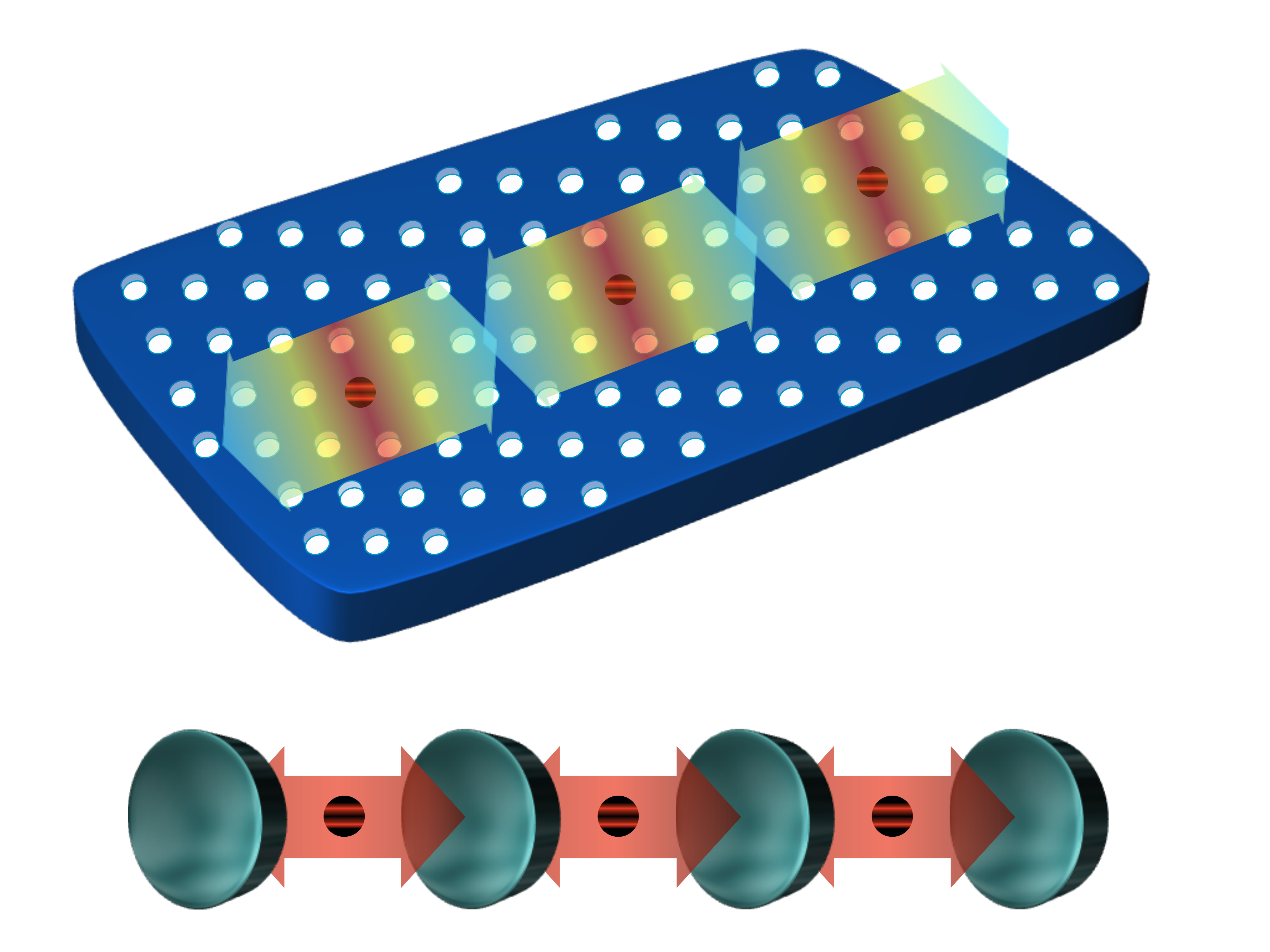}
\caption{Coupled cavities doped with two-level systems. The figure shows one possible realization of this system, in this case in photonic crystals, where cavities are defects in the periodic structure of the crystal. The two-level systems can be excitons in quantum dots or electronic levels of dopant atoms, for example.}\label{figure1}
\end{figure}

The system studied here features a chain of sites each of which containing composite excitations, also known as polaritons, created by the interaction of a boson and a fermion. A typical experimental proposal for these systems is devised in resonant cavities, the bosons being the photons that occupy the cavity mode and the fermions being  two-level electronic transitions of the onsite dopants, as depicted in figure~(\ref{figure1}) and described in~\cite{Bose,Pola}. The Hamiltonian of these coupled doped cavities (with two level impurities) is the so called Jaynes-Cummings-Hubbard model and it is given by

\begin{equation}
\mathcal{H}=\sum_{\langle i,j\rangle}^n [\mathcal{S}_i+\mathcal{T}_{(i,j)}] ,
\label{H_1}
\end{equation}
with $n$ being the number of sites, and $\mathcal{S}_i$ being the intra site Jaynes-Cummings interaction between the dopant and the resonator
\begin{equation}
\mathcal{S}_{i}= \omega_i a^{\dagger}_ia_i+\nu_i \sigma^{\dagger}_i\sigma_i+g_i(\sigma^{\dagger}_ia_i+\sigma_ia^{\dagger}_i).
\end{equation}
The $i$th site annihilation operators are $a_i$ and $\sigma_i$ for the bosonic and fermionic species, respectively. 
In Eq. (\ref{H_1}), $\mathcal{T}_{(i,j)}$ describes the photon hoping, or tunneling, between nearest neighboring sites 
\begin{equation}
\mathcal{T}_{(i,j)}=-A_{(i,j)}[a^{\dagger}_ia_{j}+a_ia^{\dagger}_{j}].
\end{equation}
The coupling strength between the two level system and the cavity in the $i$th site is given by $g_i$, and the photon tunneling strength between nearest cavities is $A_{(i,i+1)}$. The photon frequency at the $i$th site is $\omega_i$ and $\nu_i$ is the transition frequency of the dopant of the respective site, thus we define the $i$th site detuning $\Delta_i=\nu_i-\omega_i$. 

The polaritons are eigenstates of the intra-site (Jaynes-Cummings) hamiltonian and are given by $|n+\rangle=\sin(\theta_n)|g\rangle|n\rangle+\cos(\theta_n)|e\rangle|n-1\rangle$ and $|n-\rangle=\cos(\theta_n)|g\rangle|n\rangle-\sin(\theta_n)|e\rangle|n-1\rangle$, with $\tan(2\theta_n)=-g\sqrt{n}/\Delta$. The states $|n\rangle$ are photon number states and $|g\rangle$ and $|e\rangle$ are the ground and excited states of the dopant inside the cavity. Finally, the number of particles operator in the $i$th site is given by $N_i=a^{\dagger}_{i}a_{i}+\sigma^{\dagger}_{i}\sigma_i$.

\subsection{Behavior of Small Sample Systems}

Recent works show that the system described in the last session undergoes a Mott-superfluid phase transition when going from small hoping to large hoping or from negative detuning to positive detuning~\cite{Bose,JCHM}. In the first case the transition is induced because the hoping strength circumvents the photon blockage regime (non-linearity due to the dopant-cavity interaction) and in the second case the excitations are directly driven from mainly electronic (electrons are not able to hop) to mainly photonic, hence the fluidity. This phase transition can be witnessed by single site properties. For example, when the system is isolated and the average occupation number per site $\langle N_i \rangle$ is one (same number of excitations and sites), the variance of $N_i$ for any given site is a good order parameter~\cite{Bose,EN}: in the Mott phase each site has a single particle and there is no number fluctuation whereas in the superfluid phase the onsite number variance is maximum. This analysis begins to fail when one takes into account  interactions with the environment and spatial fluctuations that may not preserve the total number of particles in the system. For instance, dissipation introduces variances of the occupation number in each site and $\mathrm{var}(N_i)$ may overestimate fluidity. Furthermore, although the measure of $\mathrm{var}(N_i)$ hints at the type of excitation that  dominates each phase it cannot reveal this fundamental property in detail because it does not fully distinguish between photons, electrons and polaritons. 

We proceed to show that the entanglement between different constituents
%, a bona fide non-local measurement (local meaning on site measurement), 
not only reproduces previous results 
%with better resolution 
but actually allows for the identification of a new crossover that the previous analysis did not reveal. 

%Consider a brief example of two harmonic oscillators with the states $\sum_n^N c_n |n,N-n\rangle_{AB}$ ($N\rightarrow\infty$) and $|\alpha\rangle_A|\alpha\rangle_B$ (which could be a mean field approximation of the previous state) with $|\alpha\rangle$ being a coherent state. The site variance is not able to distinguish between these states, such that the $c_n$ coefficients might be distributed in such a way so as the states would present the same variance with the first being entangled (and could be a superfluid state) and the other being factorable, and thus, the number variance would classify both states as superfluid. 
%However, if the cavities are classically driven by external agents the number variance would increase and such increase would have nothing to do with the site-site interaction nor superfluidity, and still, entanglement would be a good indicator of fluidity.
%Furthermore, if the system is in equilibrium with a thermal environment variance is clearly not a good indicator of fluidity, since again the thermal fluctuation might increase the number variance. 

\begin{figure}[h]
\includegraphics[width=8cm]{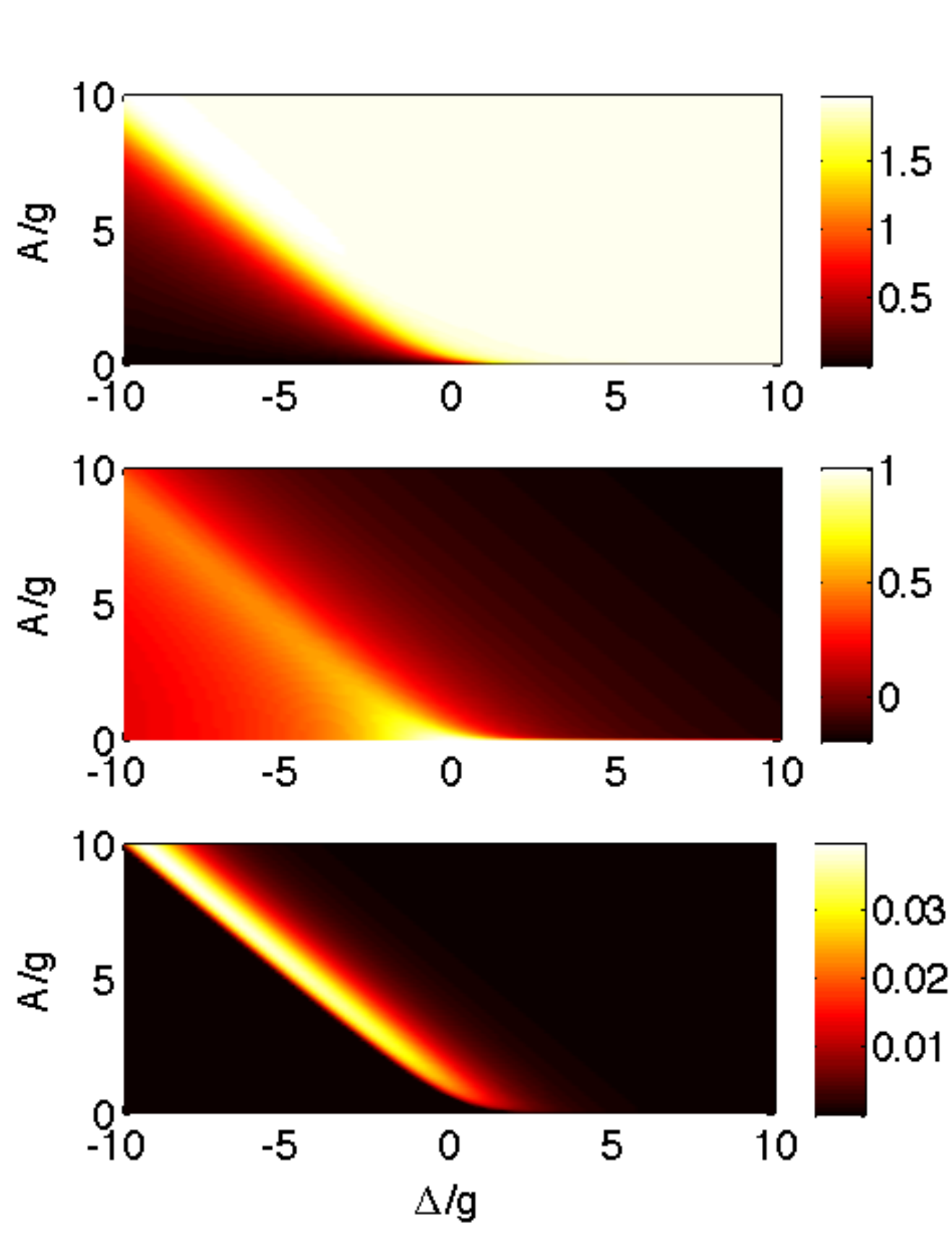}
\caption{Results for a two site system. (Top-a) Entanglement between the sites. (Middle-b) In-site Entanglement. (Bottom-c) Atom-atom entanglement.}\label{2cavitiesphase}
\end{figure}

In order to quantify the entanglement between the various components of the system we choose the negativity measure~\cite{Nega}, which is very convenient to calculate. It should be reminded that null negativity does not necessarily imply null entanglement, in fact, null negativity means null or bound entanglement. However, any non zero value of negativity guarantees some form of distillable entanglement in the system what will prove to be enough for distinguishing the different quantum phases of the system as a whole. Given the quantum state of any two constituents A and B of the system, their negativity can be found by partially transposing their reduced density matrix $R=\rho^{\mathrm{T_A}}_{\mathrm{AB}}$  and then summing up the moduli of the negative eigenvalues of $R$.

We begin with the case of small systems, where the diagonalization of the Hamiltonian is computable, by looking at properties of the lowest energy state $|G\rangle$ with the constraint of having equal number of excitations and sites $\mathcal{H}|G\rangle=E_n|G\rangle$. In other words, $|G\rangle$ is the lowest energy eigenstate of the Hamiltonian having $n=\langle \sum_i^nN_i\rangle$, with $N_i$ the number operator at the $i$th site. Consider, now, a cluster of two sites, that is the smallest possible such system. Even for this very basic \textit{unit cell}, the entanglement between the sites clearly presents the signatures of Mott and superfluid phases that were found for much larger systems in previous works. In the Mott phase (with one polariton per site) there is no entanglement between sites with the Mott insulating state being $|G(\mathrm{MI})\rangle=|1-\rangle|1-\rangle$ (for two sites). In the superfluid phase and when the excitations become mainly photonic the sites become entangled, with the superfluid state for two sites (described in~\cite{EN}) given by $|G(\mathrm{SF})\rangle=|g\rangle|g\rangle\left[\frac{1}{\sqrt{2}}|11\rangle - \frac{1}{2}(|20\rangle+|02\rangle)\right]$. It should be kept in mind that for the finite system analysis there is no phase transition, only a smoother crossover, even though the phase transition terminology is commonly adopted. The behavior of the system (phase-like diagram), quantified by the entanglement between different constituents, is depicted in figure (\ref{2cavitiesphase}), where we show the entanglement between sites, the in-site entanglement and the entanglement between atoms.

The site-site entanglement shows the phase crossover as partially presented in~\cite{ElenGround}. When the site-site entanglement is negligible the system resembles a Mott-insulator and when the site-site entanglement is non-negligible the system presents superfluid-like behavior. Thus the site-site entanglement indicates the regimes in which the system is insulating and superfluid with small and large values of entanglement, respectively [figure (\ref{2cavitiesphase},a)].
In order to quantify the polaritonic behavior we can look at the in-site entanglement that measures how correlated are the photonic field and the electronic transition in a given cavity (or a site)[figure (\ref{2cavitiesphase},b)].
In the Mott-like regime (small site-site entanglement) the in-site entanglement is significant,  indicating that the system presents polaritonic behavior. Deep in the superfluid-like regime (large site-site entanglement) the in-site entanglement is small, indicating predominant photonic behavior. However, during the crossover (as a function of either $A$ or $\Delta$) entanglement presents a non-monotonic behavior, with a region where it is maximum. Such non-monotonic increase, which is even more pronounced in the atom-atom entanglement [figure (\ref{2cavitiesphase},c)], suggests that as the system size increases and reaches the thermodynamic limit a phase transition should be verifiable, i.e. since at the point of the phase transition there are fluctuations over all length scales, more degrees of freedom interact with each other such that entanglement can exist between more degrees of freedom.
Furthermore, the regime in which in-site and site-site entanglement coexist corresponds to a polaritonic-superfluid, rather then just a photonic-superfluid.

\subsection{Large Sample Systems and the polariton-photon crossover}

We can also obtain a wider view of the system phase diagram varying the number of polaritons in the system. In order to do that we can couple the system to a chemical reservoir of polaritonic particles with chemical potential $\mu$ (at zero temperature), such that the system is in equilibrium with this reservoir. The chemical potential can be explicitly included in the system Hamiltonian $\mathcal{H}\rightarrow\mathcal{H}-\mu\sum_iN_i$. 
\begin{figure}[h]
\includegraphics[width=8cm]{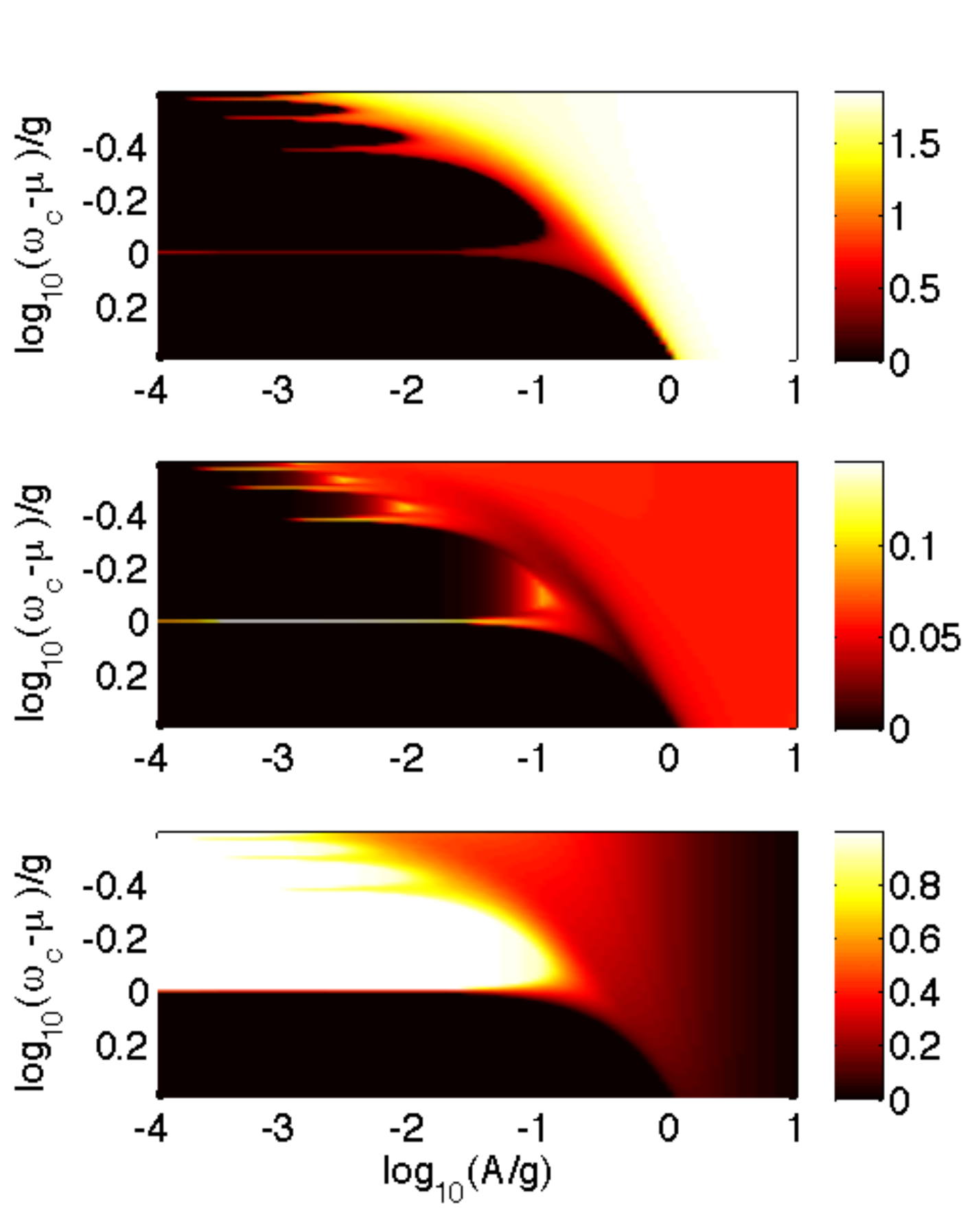}
\caption{(Top) The Mean-field parameter. (Middle) Entanglement between the sites in the cluster. (Bottom)The in-site entanglement, that is, the entanglement between the dopant and the oscillator mode in mean field theory. All data with a four-sites cluster with the dimension of each oscillator truncated to 6 photons and zero detuning ($\Delta=0$).}
\label{diagramaMu}
\end{figure}

For large (infinite) systems we adopt the mean-field approach, in which we treat a small cluster of sites interacting with a mean field, that is, a classical approximation of the rest of the chain (to which we refer to as an environment). This approach gives a factorable approximation of the non-factorable tunneling (or hoping) term by approximating the operators for their mean values plus a small fluctuation (in this case a quantum fluctuation) $a=\langle a\rangle +\delta a$. The mean field hamiltonian for the cluster becomes 
\begin{multline}\mathcal{H}_{\mathrm{MF}}=\sum_{\langle i,j\rangle{(\mathrm{cluster})}}[ \mathcal{S}_i+ \mathcal{T}_{(i,j)}]
\\+\sum_{\langle i (\mathrm{cluster}),j (\mathrm{environment})\rangle}-A_{(i,j)}[\alpha_j^{\ast}a_i+\alpha_ja_i^{\dagger}-|\alpha_j|^2],\end{multline}
with $\langle a\rangle=\alpha$ being the mean-field order parameter that has to be self-consistently determined by minimizing the ground state energy. The phase diagram as a function of the chemical potential and the hoping frequency is shown in figure (\ref{diagramaMu}) for large systems.
 
Varying the chemical reservoir we can see the Mott lobes (each lobe corresponding to plateaus of different integer numbers of polaritons) in the infinite system [figure (\ref{diagramaMu})]. The mean field parameter is null in the Mott phase and is positive in the superfluid phase. 
Only in this case we compute a site purity (one minus the purity more precisely) as the estimate of the entanglement between such site and the rest of the chain (in this case, the cluster).
Although this entanglement is not strictly zero in all of the Mott phase it still gives a fair account of the phase diagram and the lobe structure. We have considered a four-sites cluster, which is a rather small cluster, even though it already requires a considerable computational effort. Larger clusters would increase the precision of the site-cluster entanglement. The site-cluster entanglement is maximum in the lobe borders (middle panel of fig. (\ref{diagramaMu})), which, again, indicates strong polaritonic fluidity in the vicinity of the phase transition.
 
Now, looking at the in-site entanglement (bottom of fig. (\ref{diagramaMu})), which can be regarded as the very essence of the polaritons, we can see the whole picture with the overlay of the Mott-Superfluid and Hybrid-Boson crossover. The highest in-site entanglement is in the Mott-lobes and the lobe structure can also be defined by this quantity. Outside the lobes fluidity sets in, however the in-site entanglement is still very high indicating that the system has not yet undergone the Hybrid-Boson crossover despite having changed from insulating to superfluid. Farther away from the lobes and deeper into the fluid phase we finally observe the in-site entanglement vanishing indicating that the system finally turns bosonic.

\section{Disordered Small Quantum systems}

Every physical system presents imperfections  (disorder), that is, the system parameters may vary from site to site. There are many possible origins of disorder, for instance, imprecisions in the system manufacturing process, thermal fluctuations, and fluctuations induced by other uncontrollable electromagnetic sources in the system environment. One way to study the effect of disorder is to describe the parameters of each site as a stochastic variable $\xi_i$ and the Hamiltonian becomes dependent on the stochastic parameters 
$\mathcal{H}\{ \xi_i\}$. 
Naturally, the system ground state becomes dependent on the values assumed by the system parameters $|G\rangle\rightarrow|G(\{ \xi_i\})\rangle$ and there emerges a new state, an average state
\begin{equation}\rho=\int dp(\{\xi_i\})|G(\{ \xi_i\})\rangle\langle G(\{ \xi_i\})|,\label{rhostatic}\end{equation} 
that contains the statistics of the effects induced by the static disorder, with $dp(\{\xi_i\})$ being the distribution measure of the disorder, such that it gives all moments of the site parameters $\overline{\xi_i^k}=\int dp(\{ \xi_i\})\xi_i^k$. We choose to analyze only uncorrelated disorder such that the global measure is a product of local measures $dp(\{ \xi_i\})=\prod_i P(\xi_i)d\xi_i$, with gaussian distributions $P(\xi_i)=\frac{1}{\sqrt{2\pi}\delta}\exp\{-\frac{(\xi_i-\overline{\xi_i})^2}{2\delta^2}\}$. The magnitude of disorder is then given by the distribution width or the mean square deviation $\delta$.

\begin{figure}[h]
\includegraphics[width=8cm]{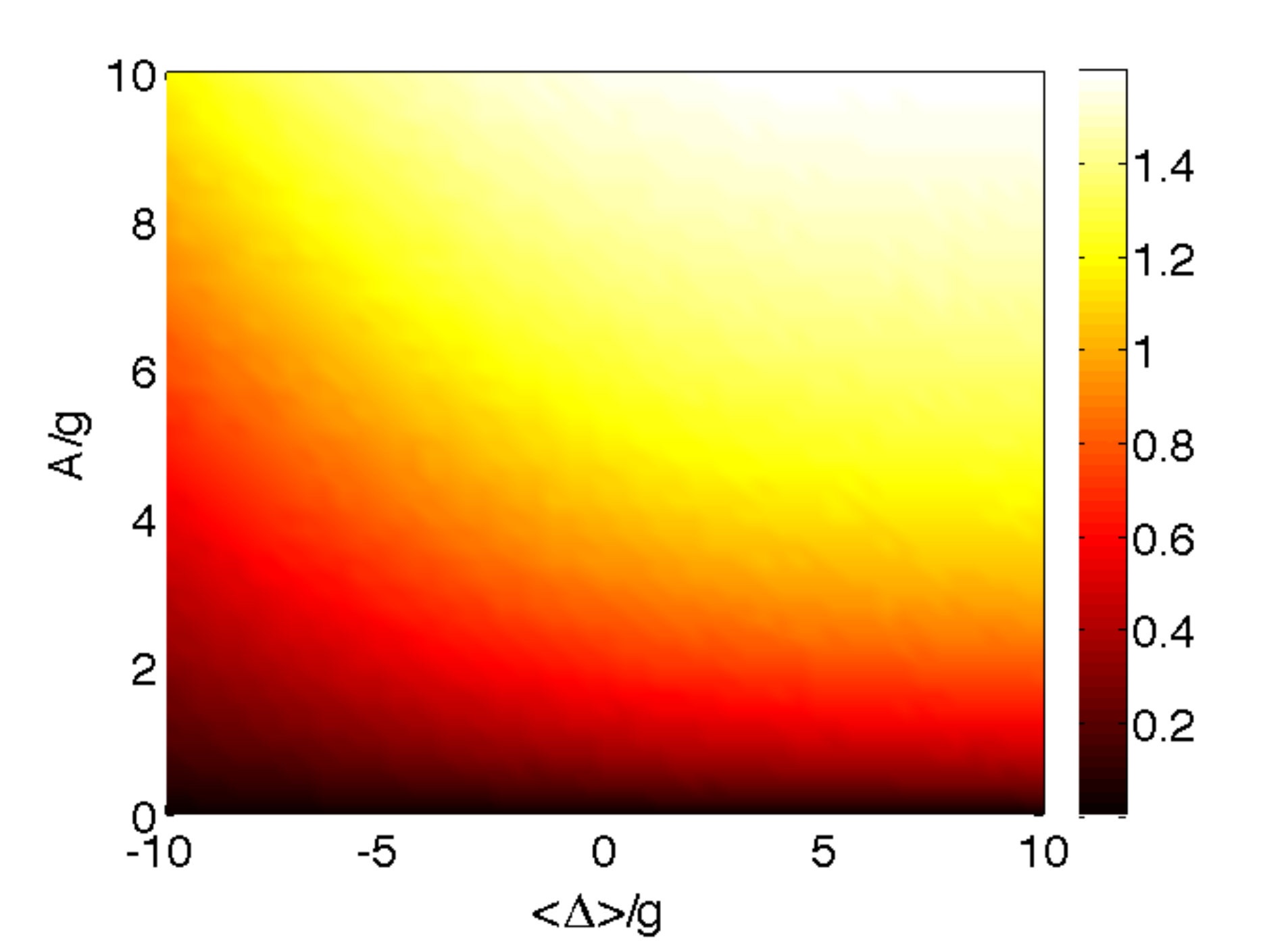}
\caption{The ensemble average entanglement $\overline{E}[\rho]$ between the two sites under disorder in the matter light detuning $\Delta$ (more specifically in the light frequency) on the phase diagram with $\delta(\Delta)=10g$.}\label{2Cavdisorder10emDelta}
\end{figure}

We can then characterize the average properties of the system given that it presents disorder. For instance we can look at the entanglement description of the phase diagram, only now we average the entanglement over the pure state ensemble generated by the different values assumed by the system parameters. We remark that the pure state ensemble given in equation~($\ref{rhostatic}$) is a physically realizable ensemble~\cite{PhyRealEns}, since in principle the experimentalist can perform spectroscopic measurements and obtain the values assumed by the parameters in that particular sample system and then prepare the system ground state. We can define the reduced states $\rho_{\mathrm{AB}}(\{ \xi_i\})=\mathrm{tr}_E\{|G(\{ \xi_i\})\rangle\langle G(\{ \xi_i\})|\}$, with the trace being performed over the environment of A and B. For instance, if we are looking at the atom-atom entanglement then we trace out the field, so the field would be the environment in this case. Therefore, we can define the average entanglement between any constituents A and B
\begin{equation}\overline{E}[\rho_{\mathrm{AB}}]=\int dp(\{\xi_i\})E[\rho_{\mathrm{AB}}(\{ \xi_i\})],\label{ArithE}\end{equation}
which is physically realizable since the ensemble of ground states also is~\cite{Nha,Phys.RealEnt}. 
Notice that the average entanglement of the ensemble is in general different than the entanglement of the average state with the usual hierarchy $\overline{E}[\rho_{\mathrm{AB}}]\geq E[\rho_{\mathrm{AB}}]$, with the average state $\rho_{\mathrm{AB}}=\int dp(\{ \xi_i\})\rho_{\mathrm{AB}}(\{ \xi_i\})$.

\begin{figure}[h]
\includegraphics[width=8cm]{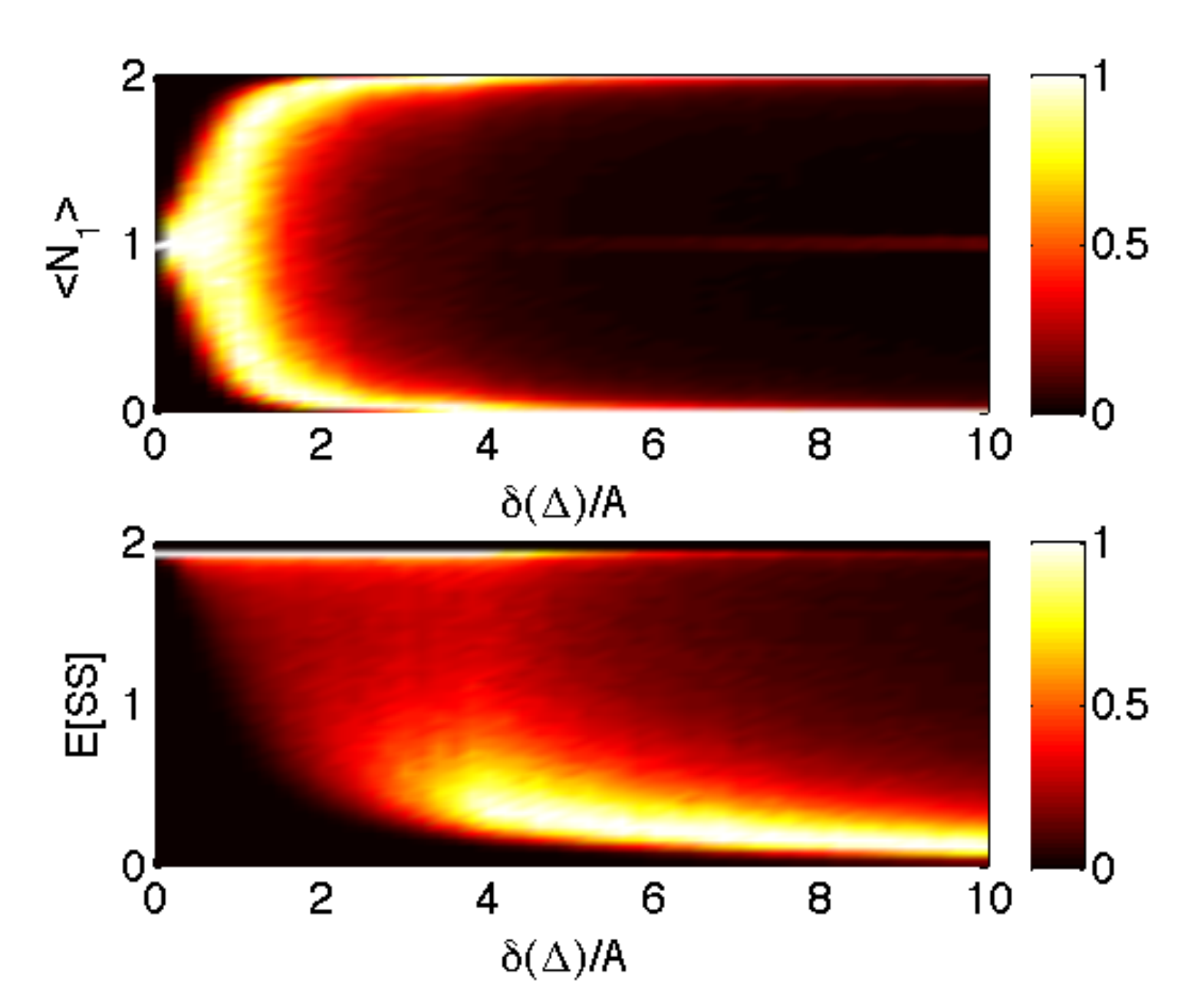}
\caption{ Probability distributions induced by disorder in the matter light detuning. (Top) Renormalized histogram $P(\langle N_1\rangle)$ of the site occupation number as a function of disorder. (Bottom) Renormalized histogram $P(E)$ of the site-site entanglement as a function of disorder. The light hoping is $A=g$ and average detuning $\langle\Delta\rangle=5g$.}\label{HistoA1Delta5}
\end{figure}

In what follows in this section we consider only two sites of the jaynes-Cummings-Hubbard hamiltonian. 

\subsection{Disorder in the matter-light detuning}

Now we can describe the effects induced by disorder in each of the parameters individually, detuning $\Delta$, hoping $A$, and matter light coupling $g$. 
Let us begin by analyzing disorder only in the cavity-atom detuning, thus
$\{\xi_i\}=\{\Delta_i\}$ (see figures (\ref{2Cavdisorder10emDelta}) and
(\ref{HistoA1Delta5})). As can be seen in fig. (\ref{2Cavdisorder10emDelta})
the average entanglement between sites seems to decrease over the whole phase
diagram in comparison with the clean case of fig. (\ref{2cavitiesphase}a). 

\begin{figure}[h]
\includegraphics[width=8cm]{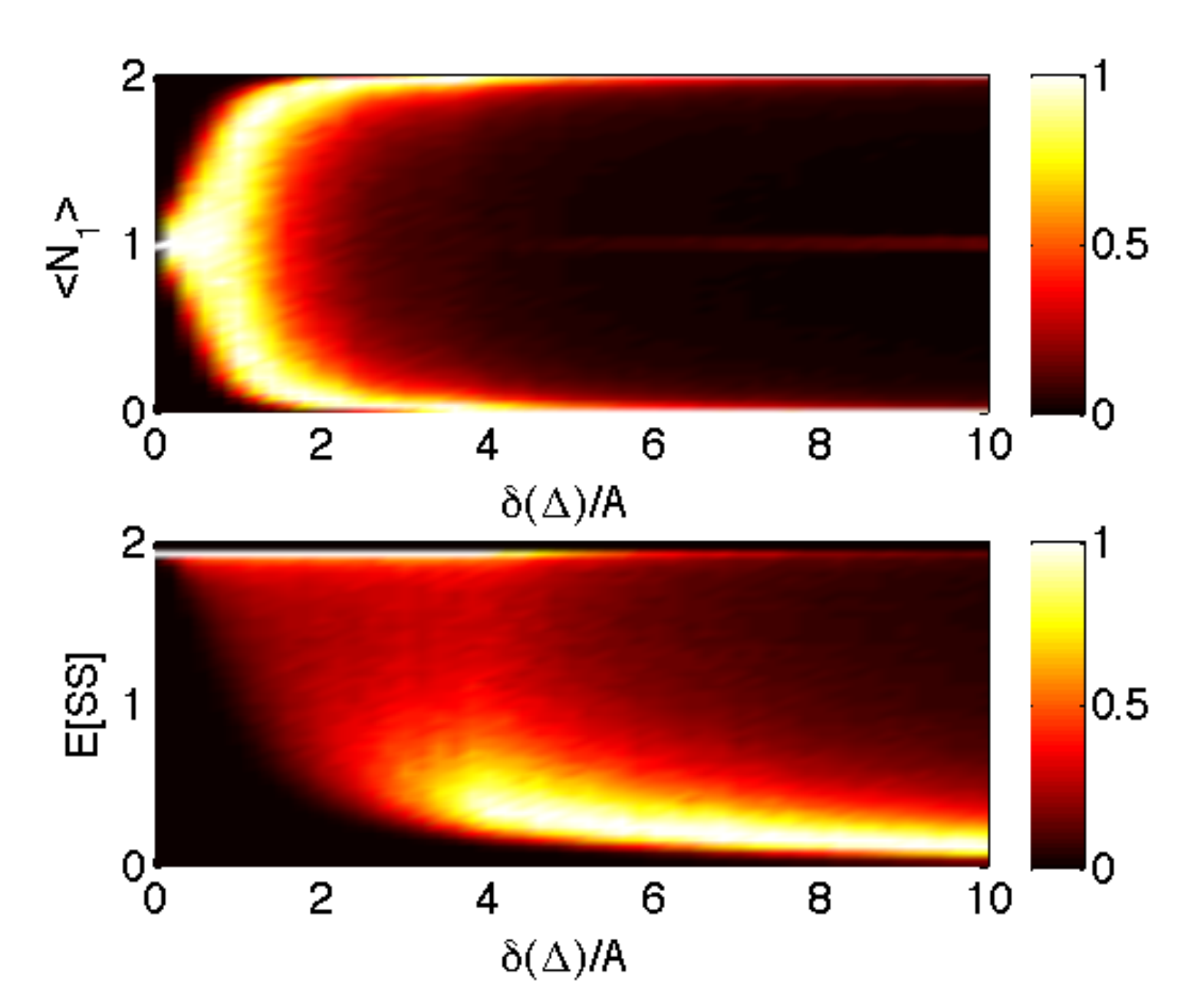}
\caption{Effects of static disorder in the light hoping $A$ on the phase diagram. The ensemble average entanglement $\overline{E}[\rho]$ between the two sites. Disorder of $\delta(A)=10g$.}\label{2Cavdisorder5emA}
\end{figure}

The decrease of site-site entanglement indicates that the excitations tend to localize through an Anderson-like mechanism. For instance,
starting with the system in the superfluid phase as we
increase the detuning disorder the distribution of the single site number
occupation $P(\langle N_1\rangle)$ is broadened and then it becomes a two
peaked distribution (see top panel of figure (\ref{HistoA1Delta5})). In the regime in which
the distribution $P(\langle N_1\rangle)$ presents two peaks the system is
fully localized, such that one of the peaks corresponds to all excitations in
cavity one and the other corresponds to zero excitations in the cavity. This
extreme regime of localization can be regarded as a bosonic
bunching: the large disorder in the cavity line width allows
for realizations in which the cavity has a very low frequency such that it is
energetically favorable to fit more than one excitation in one cavity instead
of distributing the excitations over the sites. For two sites
and two excitations the state can be approximately given by
$|2\rangle|g\rangle|0\rangle|g\rangle$ or
$|0\rangle|g\rangle|2\rangle|g\rangle$ with the atoms in their ground states.

The parameters in fig. (5) are such that the system is in the superfluid phase
in the clean limit. In this case, as expected,
the distribution of the site-site entanglement $P(E[\mathrm{SS}])$ shows a
peak at the maximum value (bottom panel of the figure). However as disorder increases there emerges a second peak close to
the minimum value corresponding to localized states. Thus the
ensemble presents both superfluid and insulating {\it states}
for intermediate values of disorder. The presence of the two sorts of states
can be regarded as a precursor of a glassy phase~\cite{Small1}, which we will
show to be true with the large system analysis. Furthermore, as disorder
increases even further the system becomes fully localized and the site-site entanglement distribution becomes single-peaked at very small values of entanglement. 

\begin{figure}[h]
\includegraphics[width=8cm]{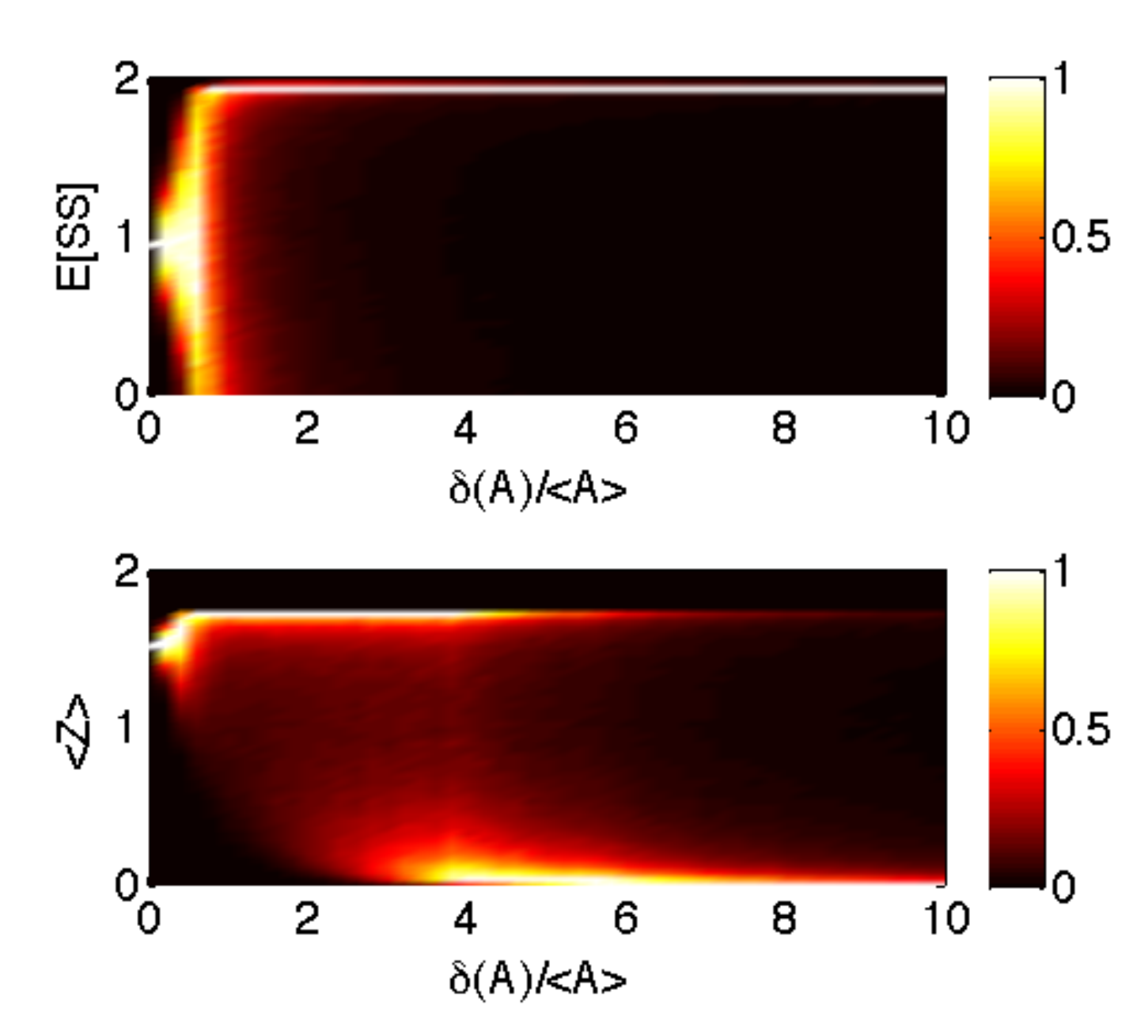}
\caption{Probability distributions induced by disorder in the light hoping. (Top) Renormalized histogram $P(\langle Z\rangle)$ of the total atomic occupation number as a function of disorder. (Bottom) Renormalized histogram $P(E)$ of the site-site entanglement as a function of disorder. The average light hoping is $\langle A\rangle=1g$ and detuning $\Delta=-2g$.}\label{HistoA1Delta-5emA}
\end{figure}

\subsection{Disorder in the photon hopping}

Disorder in the photon hoping generates a different effect (see figures
(\ref{2Cavdisorder5emA}) and \ref{HistoA1Delta-5emA}). Carrying out the same
analysis as for disorder in the detuning, we see that the
average site-site entanglement increases in the region where a Mott
 phase exists in the clean limit, while it remains practically unaltered in
the superfluid phase. Fluctuations in the hoping may actually induce a glassy
fluid phase~\cite{Small,DIfluid,SMFT2}, that is, disorder allows realizations
in which the hoping is stronger than the photon blockade and those
realizations may prevail.
The $A$-disorder may also suppresse the polaritonic behavior which can be
seen as a decrease in the atomic population. We can also look at the
distribution for the total atomic excitation
$Z=\sum_i\sigma^{\dagger}_i\sigma_i$ as a function of disorder
starting from the system at the Mott phase in the clean limit. The distribution $P(\langle Z\rangle)$ is very asymmetric and as disorder increases it concentrates at the extreme values assumed by $\langle Z\rangle$. One of the extremes corresponds to polaritonic superfluidity and the other to photonic superfluidity, with the latter prevailing in the limit of very large disorder. This can be corroborated by the entanglement distribution $P(E[\mathrm{SS}])$.

 \subsection{Disorder in the matter-light coupling} 
 
\begin{figure}[h]
\includegraphics[width=8cm]{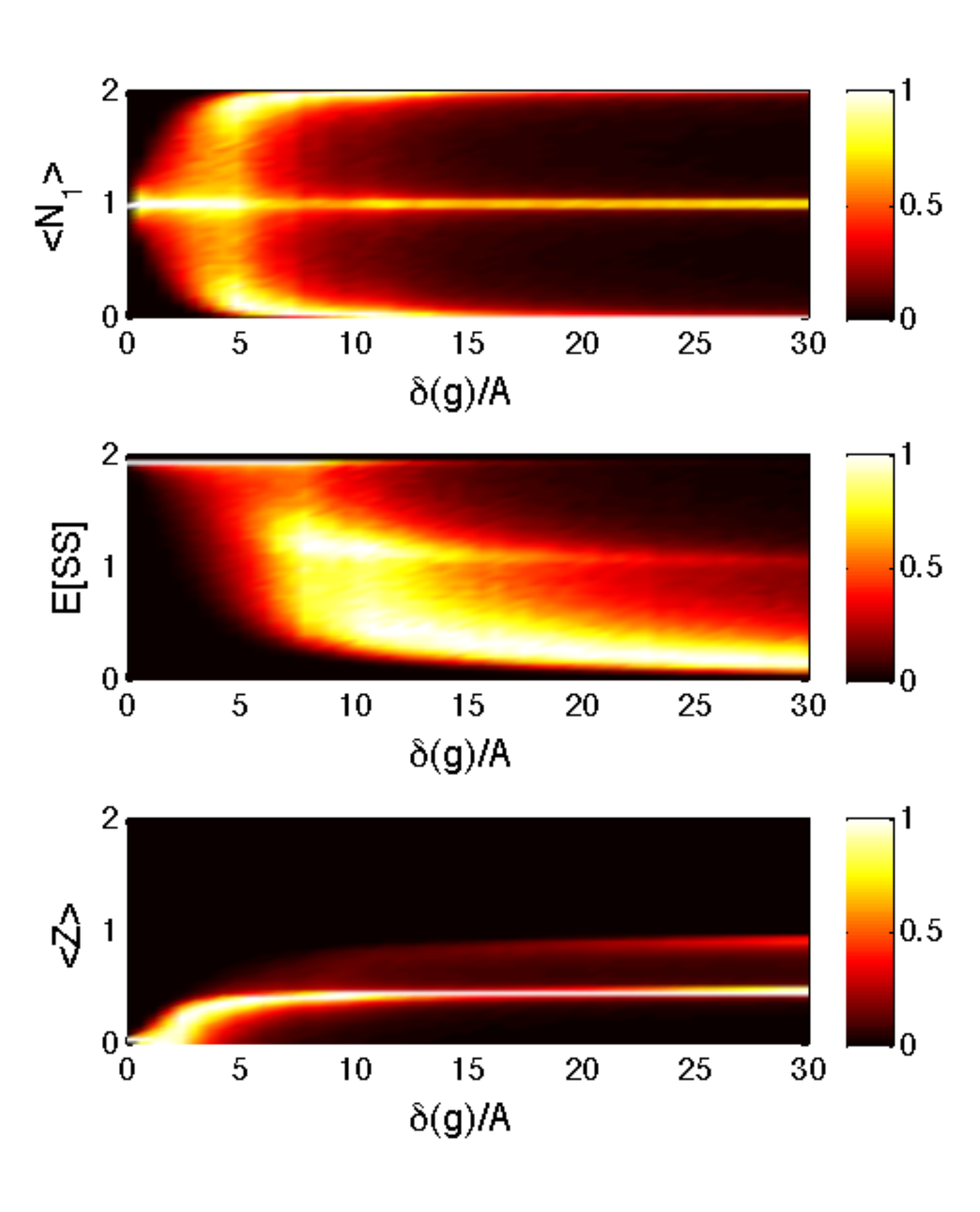}
\caption{Probability distributions induced by disorder in the matter-light coupling $g$.
 (Top) Renormalized histogram $P(\langle N_1\rangle)$ of the site occupation number as a function of disorder. (Middle) Renormalized histogram $P(E)$ of the site-site entanglement as a function of disorder. (Bottom) Renormalized histogram $P(\langle Z\rangle)$ of the total atomic occupation number as a function of disorder. The light hoping is $A=1g$ and detuning $\Delta=5g$.}\label{HistoA1Delta5emg}
\end{figure}

 Finally we analyze disorder in the matter light coupling $g$ (see figure
 (\ref{HistoA1Delta5emg})). A first look at the $g$-disordered average
 entanglement (not shown here since it resembles very closely the
 $\Delta$-disorder case) suggests that disorder in the Jaynes-Cummings
 coupling also induces localization, that is, disorder allows a great number of
 meaningful realizations in which the sites are almost unentangled and the
 excitations tend to bunch. However, the $g$-disorder induced distribution of
 the site occupation number $P(\langle N_1\rangle)$ (top panel of
   fig. (\ref{HistoA1Delta5emg})) is very different than the one induced by
 $\Delta$-disorder (which we showed that presents localization). In
 the current case the $P(\langle N_1\rangle)$ distribution
 shows three peaks, the extreme ones corresponding to bunching similar to the
 $\Delta$-disorder case, and a middle one that corresponds to states in which
 the excitations are still equally distributed among the sites. Disorder in
 the matter-light coupling induces states with Mott-like features in which the
 site-site entanglement vanishes (see middle panel of the
   figure). In fact, this is the meaning of the middle peak in $P(\langle
   N_1\rangle)$: some sites undergo a superfluid-insulating transition through a Mott-like mechanism. Since the distribution of the atomic occupation $P(\langle Z\rangle)$ (bottom panel of
   fig. (\ref{HistoA1Delta5emg})) is narrowly centered at an appreciable (although not extreme) value we may conclude that the system nature becomes mainly polaritonic, and thus presenting both Mott states (middle peak in $P(\langle N_1\rangle)$ and polaritonic bunching (the extreme peaks in $P(\langle N_1\rangle)$. This suggests that the system behaves very similarly to an Anderson-Mott insulator \cite{Carol}.

The presence of superfluid, Mott and Anderson-like states for intermediate disorder in
the coupling $g$ suggests that glassy phases would be induced in larger
systems of the Jaynes-Cumming-Hubbard type. In fact, a different situation was
analyzed in~\cite{Fazio}, in which there is disorder in the number of
impurities per cavity. Since the number of atoms fluctuates then the intensity
with which light couples to matter also fluctuates, and it was shown
in~\cite{Fazio} that such disorder induces glassy phases. Interestingly, even
small versions of the quantum system present evidence of many diverse phases
and behaviors expected only for larger quantum systems, which
  we discuss in the next section.
  
\section{Disordered Large Quantum Systems}

To address the physics of large disordered quantum systems we apply a recently developed technique, namely stochastic-mean-field-theory (SMFT)~\cite{SMFT}. This method has been shown to provide appropriate descriptions for the effects of disorder without over estimating coherence and fluidity and has already been successfully applied to the disordered Bose-Hubbard model~\cite{SMFT}. The main reason for the effectiveness of the method is self-consistently determining the probability distribution for the mean-filed parameter $P(\alpha)$ (instead of $\alpha$ itself) through an iterative process. 

\subsection{Stochastic-Mean-Field-Theory}

Firstly we describe how to account for any on site disorder (with constant photon hopping $A$), afterwards we describe the special case of hopping disorder. 
In the mean-field description every site has a number $z$ of nearest neighbors. The mean-filed hamiltonian for the $k$th site depends only on the scaled sum 
\begin{equation}\eta_k=\sum_{ j }A_{\langle k,j\rangle}\alpha_j.\end{equation} 

The probability distribution for $\eta$ (we drop the site index for convenience) can be found from
a simple and fundamental relation known as the convolution theorem 
\begin{equation}Q(\eta)= \int\ldots\int \prod_i^z d\alpha_i P(\alpha_i)\delta\left(\eta-A\sum_j^z \alpha_j\right),\end{equation}
which reduces to the Fourier transforms $\varphi(\beta)=\int d\alpha P(\alpha)e^{i\beta\alpha}$ and
\begin{equation}Q(\eta)=\frac{1}{2\pi A}\int d\beta [\varphi(\beta)]^{z}e^{-i\eta\beta/A}.\label{Q} \end{equation}

The first step in the algorithm is to choose a trial distribution for $\alpha$ (different from a delta centered at $\alpha=0$) and the desired distribution (in our case a Gaussian) for the disordered parameter $\xi$ (the detuning or matter-light coupling). Then we assume all $\alpha_j$ to be independent form each other and we determine the self-consistent distribution 
\begin{equation}P(\alpha)=\int\int dq(\eta) dp( \xi) \delta(\alpha-\langle a\rangle),\end{equation}
such that $\langle a\rangle=\langle G[\xi,\eta]|a|G[\xi,\eta]\rangle$, with $dq(\eta)= d\eta Q(\eta)$. The procedure is iterated until we observe convergence, that is, until $P^{(i)}(\alpha)$ in the $i$th step is statistically close to $P^{(i+1)}(\alpha)$. Finally, the average state of the site is given by the disorder-induced ensemble
\begin{equation}\rho=\int\int dq(\eta) dp(\xi) |G[\xi,\eta]\rangle\langle G[\xi,\eta]|.\end{equation}

To account for disorder in the photon hopping we must add another step to the procedure. It is convenient  
to work with the variable $\phi=A\alpha$ whose probability distribution is given by (we use a subindex to distinguish the various distributions)
\begin{equation}P_{\phi}(\phi)=\int\int dAd\alpha P_A(A)P_{\alpha}(\alpha)\delta(A\alpha-\phi).\end{equation}
Then we can determine $Q(\eta)$ through the usual Fourier transforms $\varphi(\beta)=\int d\phi P_{\phi}(\phi)e^{i\beta\phi}$ 
\begin{equation}Q(\eta)=\frac{1}{2\pi}\int d\beta [\varphi(\beta)]^ze^{-i\eta\beta}\end{equation}
and the procedure follows as for the case of on site disorder.

\section{Disorder-induced transitions}

The results presented in this section will allow us to conclude that the asymptotic effects of disorder (very large disorder) in the thermodynamical limit are very similar to the effects in small samples of the system. However, there are some significant quantitative differences, for instance, there are in fact phase transitions induced by disorder in the thermodynamical limit. It should also be pointed out that our approach is a bit different in this section. From now on we work with single-site mean-field-theory rather than cluster-mean-field-theory, and we follow this strategy to avoid higher computational demands. This limits the applicability of the method and quantities like the site-cluster entanglement are no longer addressable. Nonetheless, we are able to increase the local effective dimension of the oscillator to 20 states. Another difference between the approaches for small and large samples is that in the first case we fix the number of excitations in the system and it does not change as disorder increases. This is not the case in the present section, and in fact, the total number of excitations may change as a function of disorder. 

Using the SMFT approach we are able to perform an analysis of the thermodynamical limit.
The method allows us to recover the probability distributions (under the single-site-mean-field approximation) for the various quantities we analyze to characterize the system. 
The average mean field parameter, for instance, can be readily evaluated as $\langle\alpha\rangle=\int \alpha P(\alpha) d\alpha$.
We follow the same ordering of presentation of the results: Firstly, we show the results for the detuning disorder, then for the hopping disorder and finally for the matter-light coupling. 
Given the unlimited nature of the disorder distribution we analyze (Gaussian) it follows that the insulating phases we present below are of glassy nature. Such phases have non vanishing number variance as opposed to the Mott-insulating phases~\cite{SMFT2}.
Therefore glassy insulators can be characterized by vanishing superfluidity and non vanishing compressibility (which can be related to the number variance). However, we do not show the compressibility of the system, since the result can be readily anticipated. The compressibility increases in the insulating regions, thus glassy phases are established.

\begin{figure}[h]
\includegraphics[width=10cm]{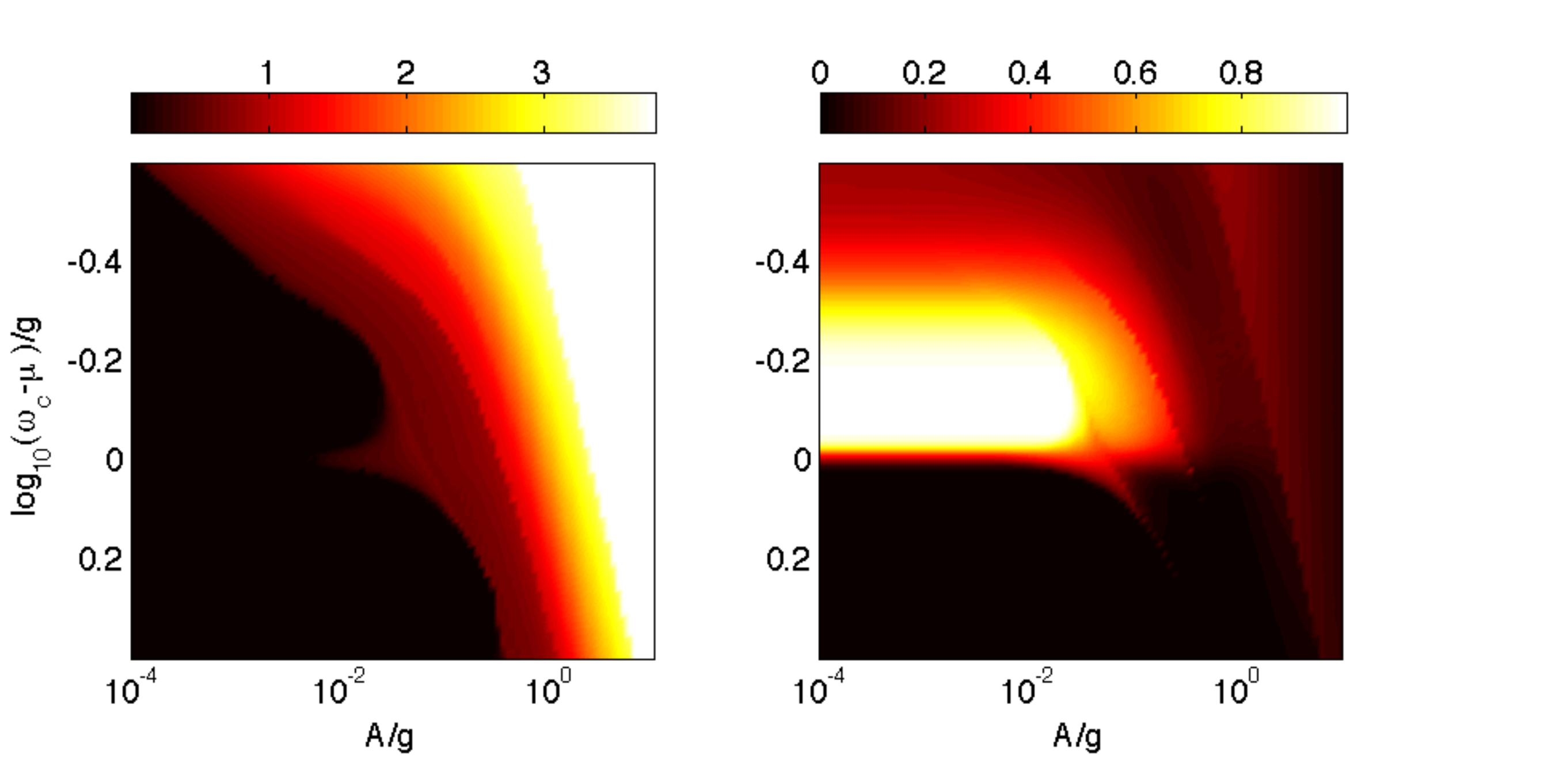}
\caption{Stochastic mean field results for disorder in the matter-light detuning with $\delta(\Delta)=0.1g$ and $\langle\Delta\rangle=0$. (Left) The average mean field parameter. (Right) The entanglement of the average state.}\label{StochDemDelta}
\end{figure}

\begin{figure}[h]
\includegraphics[width=10cm]{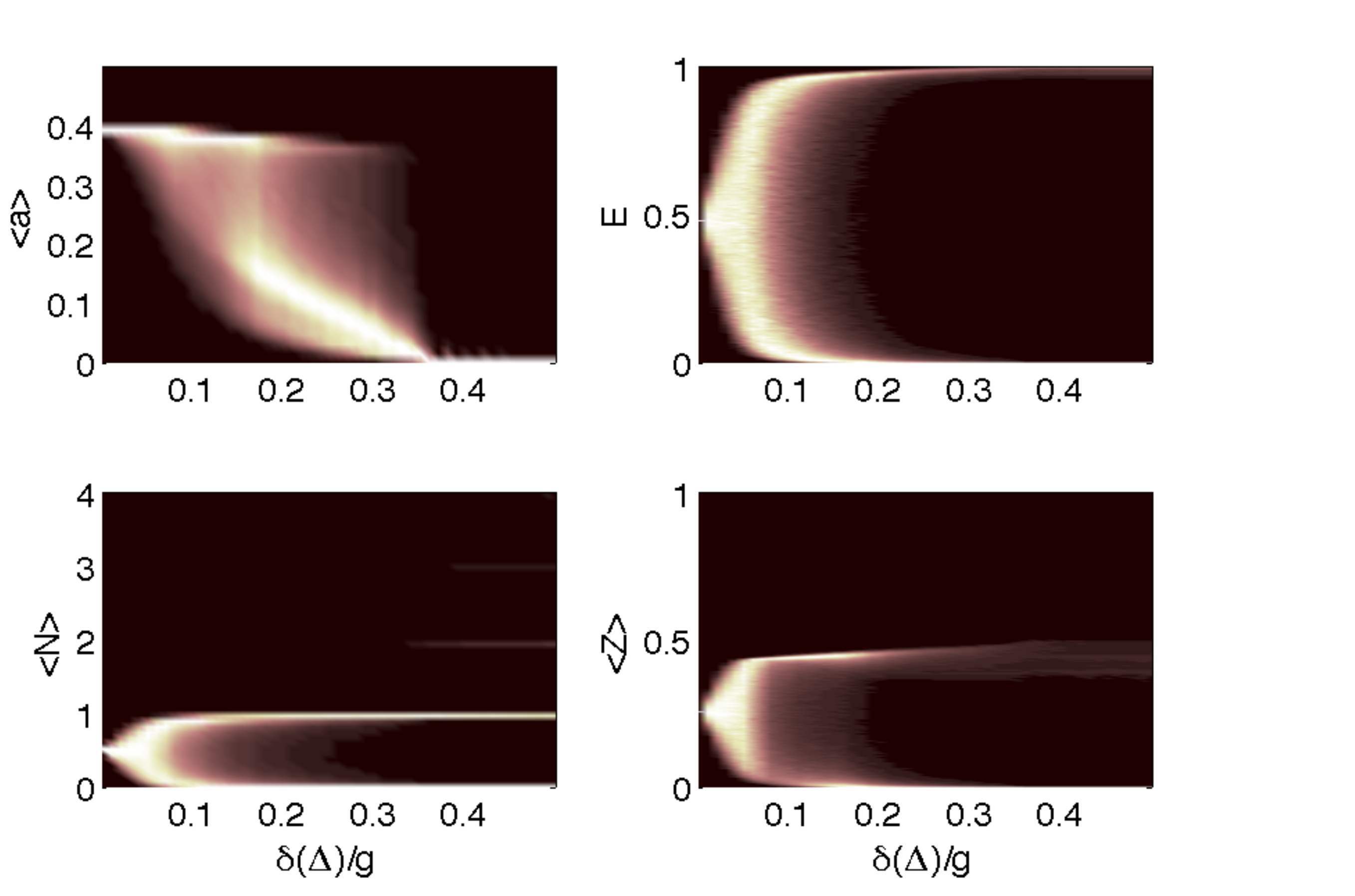}
\caption{Probability distributions as a function of disorder in the matter-light detuning with $A=10^{-1.9}g$, $(\mu-\omega_c)/g=-1$ and $\langle\Delta\rangle=0$. Black corresponds to vanishing probability. (Top-Left) Distribution for the mean field parameter, (Top-Right) for the entanglement of the average state, (Bottom-Left) for the cavity excitation and (Bottom-Right) for the atomic excitation.}\label{SMFThistemD}
\end{figure}

\subsection{Disorder in the matter-light detuning}

As show in figure (\ref{StochDemDelta}) the net effect of disorder in the detuning is to induce insulating behavior, indicated by the destruction of the fluid phase surrounding the Mott-lobes, in fact, the lobe structure disappears for significant amounts of disorder. The in-site entanglement shows that the system remains in superpositions of light and matter excitations for intermediate values of disorder. However, as we increase disorder the in-site components are either highly entangled or unentangled with higher probability. We can see the distributions as functions of disorder in figure (\ref{SMFThistemD}). The transition from fluid to insulating is evident in the distribution of the mean field parameter. 
All this corroborates the small system predictions and once again the system is mainly photonic for strong disorder. Interestingly, in the present limit the distribution of cavity excitation ($P\langle N\rangle$ in figure (\ref{SMFThistemD})) is a series of delta functions (with different weights) centered at integer values of the mean occupation, which is in agreement with the insulating and bunched behavior.

\begin{figure}[h]
\includegraphics[width=10cm]{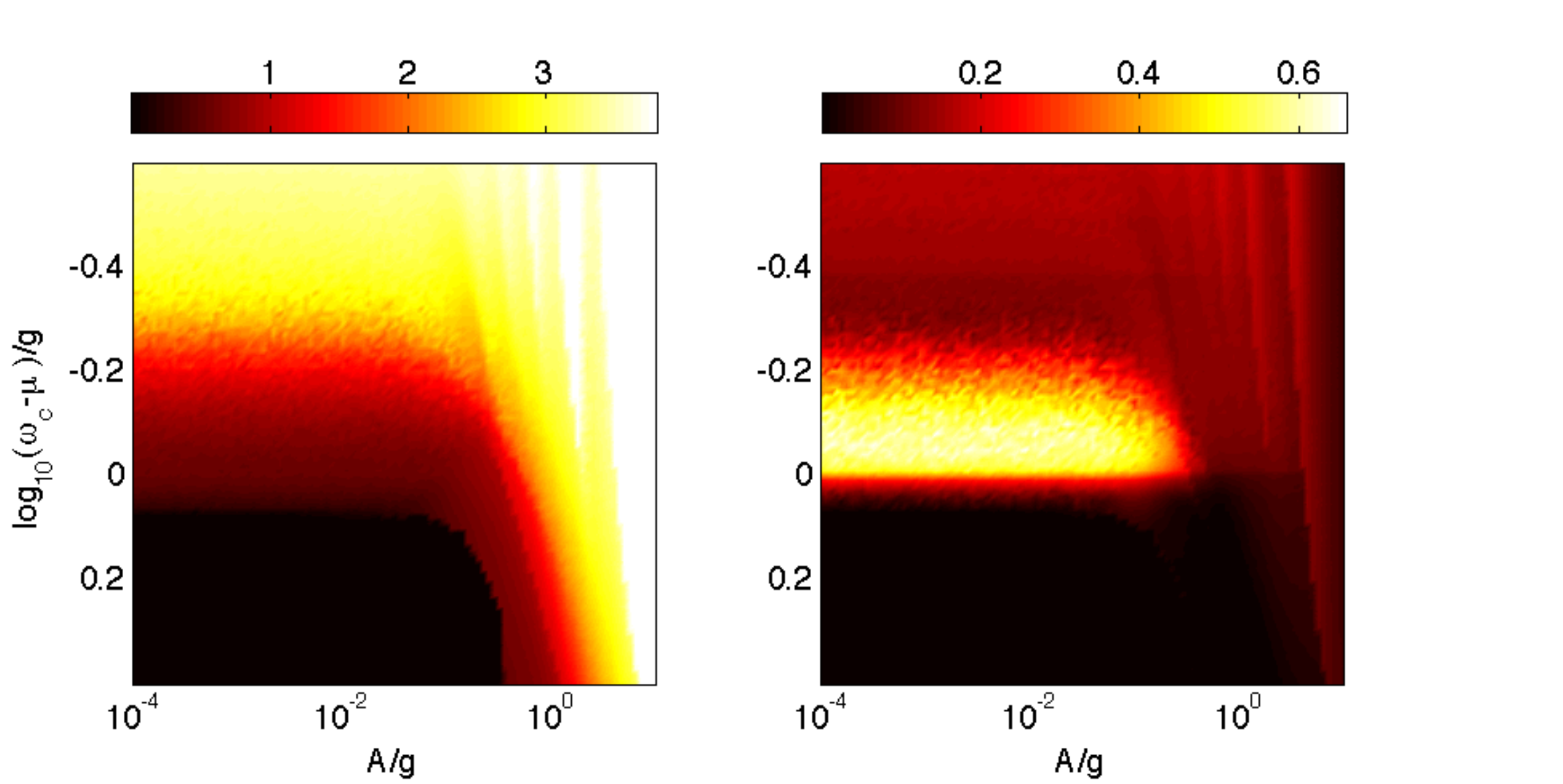}
\caption{Stochastic mean field results for disorder in the hopping with $\delta(A)=0.1g$ and $\Delta=0$. (Left) The average mean field parameter. (Right) The entanglement of the average state.}\label{StochDemA}
\end{figure}

\subsection{Disorder in the photon hopping}

The effects that hopping disorder produces in the system are opposite
to those produced by detuning disorder, as it is the case for small
systems. In the current case disorder induces induces a fluid phase, as we can see in figure (\ref{StochDemA}), and decreases the in-site entanglement indicating that the system becomes more photonic in nature. As in other situations analyzed, by looking at the distributions of
the physical quantities as a function of disorder (not shown), we were
able to observe the transition from insulating to fluid behavior as
disorder in the hopping parameter increased.

\begin{figure}[h]
\includegraphics[width=10cm]{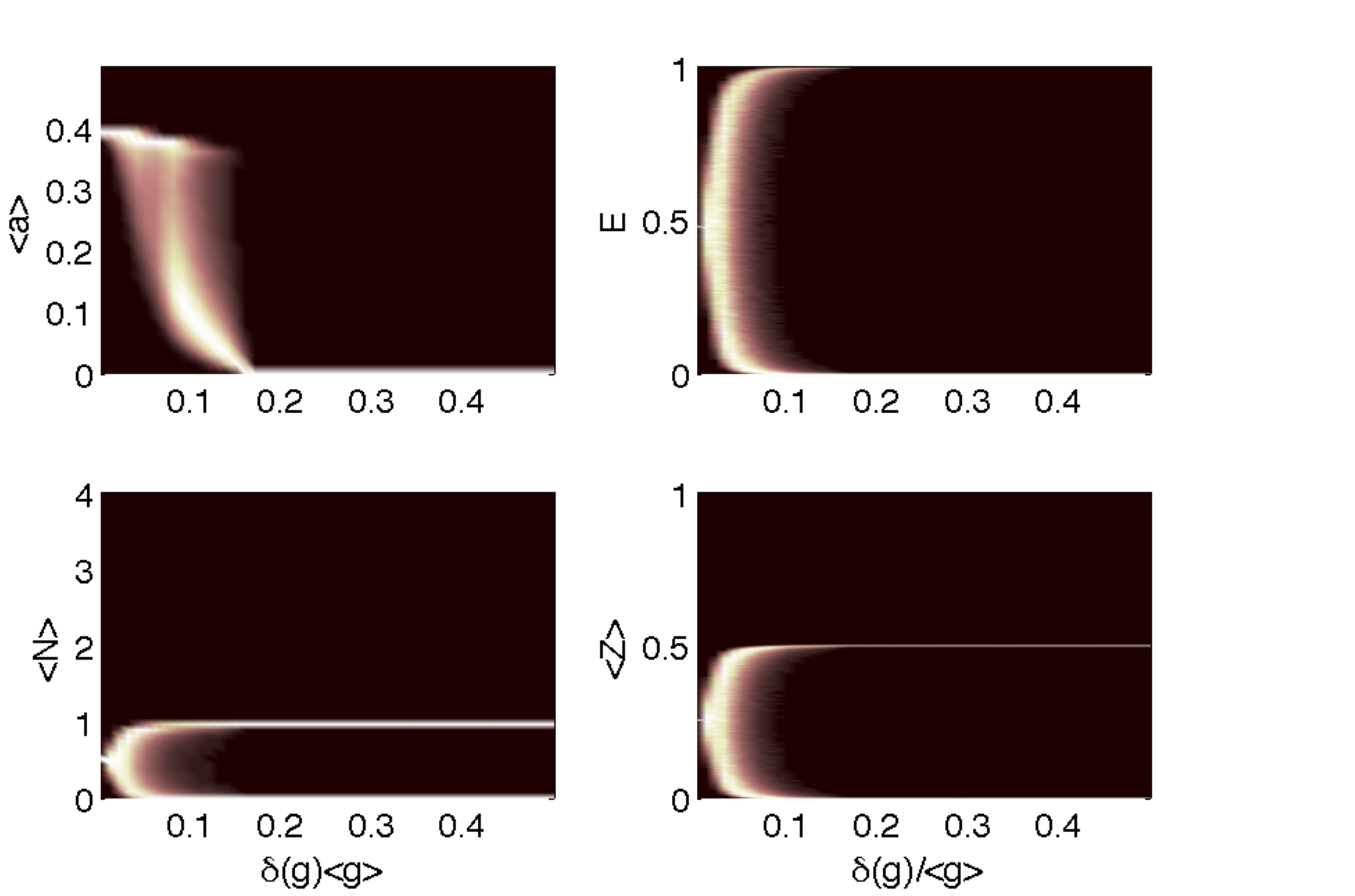}
\caption{Probability distributions as a function of disorder in the matter-light coupling with $A=10^{-1.9}g$, $(\mu-\omega_c)/g=-1$ and $\Delta=0$. Black corresponds to vanishing probability.  (Top-Left) Distribution for the mean field parameter, (Top-Right) for the entanglement of the average state, (Bottom-Left) for the cavity excitation and (Bottom-Right) for the atomic excitation.}\label{SMFThistemg}
\end{figure}

 \subsection{Disorder in the matter-light coupling}

Finally, the effects of disorder in the light-matter coupling, once again, resemble the ones induced by the detuning disorder. Even though both the detuning and coupling disorders induce insulating behavior, it should be pointed out they do it through very different physical mechanisms. 

In the detuning case cavities may be in lower frequencies in many sites which allows for more photons to localize (even several photons per cavity). In the coupling case the strength of the polaritonic nature may be increased in cavities that are strongly coupled to their corresponding matter components inducing Mott behavior. And as we can see in figure (\ref{SMFThistemg} bottom left) in comparison with fig. (\ref{SMFThistemD} bottom left) the distribution for the cavity population does not present the higher order peaks, only the ones corresponding to zero or one polariton per cavity. This behavior is due to the fact that the sites with strong matter-light coupling prevent the accumulation of larger numbers of particles per site (Mott mechanism).

Adding the information provided by the entanglement and atomic population distributions, we have found that a fraction of the sites assumes the Mott behavior and the rest are localized or even empty. Thus, we corroborate the small system analysis that suggests that the system behaves very similarly to the Anderson-Mott insulator~\cite{Carol}. 
It is, however, strikingly interesting that in one case (disorder in the detuning) it is the low frequency sites that dominate the resulting statistical behavior and in the other case (disorder in the matter-light coupling) it is the strongly coupled sites that dominate, even though the distribution of the disorder parameter is gaussian and unbiased.

\section{Conclusions}
 
 We were able to characterize the phase diagram and the Mott-Superfluid transition of small and large samples of the Jaynes-Cummings-Hubbard Hamiltonian using entanglement measures between the various possible partitions of the components of the system. In particular, we showed that these non-local measures identify more clearly where the transition happens. Furthermore, and more importantly, we also showed that entanglement measures distinguish which type of excitation dominates each phase which in turn allowed us to identify a crossover that is particular to this Hybrid system and does not have a purely bosonic analog. This behavior splits the superfluid regime into two, the first one dominated by polaritons and the second is purely photonic. 

For the disordered system we have shown that the simple statistical treatment of small systems can be quite instructive and allows us to draw conclusions that can be corroborated by the large system limit. We have shown that disorder both in the light-matter detuning and light-matter coupling induce insulating phases, however they do it through very different physical mechanisms. The former allows for photonic localization and bunching, the latter induces Mott behavior in a fraction of the sites that prevents the bunching. Furthermore, the cavity-cavity coupling disorder induces a glassy fluid phase. The rich in-site structure of the system leads to these diverse disordered phases with very different statistics and physical meanings. 
 
A great amount of work remains to be done on the characterization of the JCH system. As a valuable point we suggest that an appropriate and efficient method should be applied to the study of the large hopping limit (with and without disorder) in which the mean field approach adopted here is limited by the truncation of the state space.

Finally, it is worth mentioning once again the mesoscopic aspect of the systems proposed to implement the JCH Hamiltonian as well as the increasing ability to manipulate the different parameters of these systems, sometimes even at an individual level. These properties suggest that it will be possible to carry on a thorough experimental investigation of the effects of disorder and its relation to phase transitions and entanglement in many body physics in the near future.
 
\acknowledgements

We would like to thank U. Bissbort for correspondence and CNPq and Fapemig for financial support.


\begin{thebibliography}{C}

\bibitem{QFT}D. Jaksch, C. Bruder, J. I. Cirac, C. W. Gardiner, and P. Zoller
Phys. Rev. Lett. \textbf{81}, 3108 (1998); Markus Greiner, Olaf Mandel, Tilman Esslinger, Theodor W. H穫sch, Immanuel Bloch, Nature \textbf{415}, 39-44 (2002).
\bibitem{Bose} D.G. Angelakis, M. F. Santos, and S. Bose, Phys. Rev. A \textbf{76}, 031805(R) (2007). 
\bibitem{EN}  E. K. Irish, C. D. Ogden, M. S. Kim, Phys. Rev. A \textbf{77}, 033801 (2008).
\bibitem{ElenGround} E. K. Irish, Phys. Rev. A \textbf{80}, 043825 (2009).

\bibitem{Small1} Dirk-Soren Luhmann, Kai Bongs, Klaus Sengstock, and Daniela Pfannkuche, Phys. Rev. A \textbf{77}, 023620 (2008).
\bibitem{Small} Qi Zhou and S. Das Sarma, Phys. Rev. A \textbf{82}, 041601(R) (2010).

\bibitem{DMRGCavs} Davide Rossini, Rosario Fazio and Giuseppe Santoro, EPL \textbf{83}, 47011 (2008).


\bibitem{CriticalEnt} A. Osterloh, Luigi Amico, G. Falci, Rosario Fazio, Nature \textbf{416}, 608-610 (2002); G. Vidal, J. I. Latorre, E. Rico, and A. Kitaev, Phys. Rev. Lett. \textbf{90}, 227902 (2003).

\bibitem{OrdeEnt} Fernando Guadalupe Santos Lins Brandao, New J. Phys. \textbf{7}, 254 (2005). 	
 


\bibitem{JCHM} A. D. Greentree, C. Tahan, J. H. Cole, and L.C.L. Hollenberg, Nat. Phys. \textbf{2}, 846 (2006); M. I. Makin, Jared H. Cole, Charles Tahan, Lloyd C. L. Hollenberg, and Andrew D. Greentree, Phys. Rev. A \textbf{77}, 053819 (2008); Davide Rossini, Rosario Fazio, Giuseppe Santoro, Europhys. Lett. \textbf{83}, 47011 (2008); Jens Koch and Karyn Le Hur, Phys. Rev. A \textbf{80}, 023811 (2009); S. Schmidt and G. Blatter
Phys. Rev. Lett. \textbf{103}, 086403 (2009); Alexander Mering, Michael Fleischhauer, Peter A. Ivanov, and Kilian Singer,
Phys. Rev. A \textbf{80}, 053821 (2009); James Quach, Melissa I. Makin, Chun-Hsu Su, Andrew D. Greentree, and Lloyd C. L. Hollenberg
Phys. Rev. A \textbf{80}, 063838 (2009); S. Schmidt and G. Blatter
Phys. Rev. Lett. \textbf{104}, 216402 (2010).


\bibitem{Dario} A. Tomadin, V. Giovannetti, R. Fazio, D. Gerace, I. Carusotto, H.E. Tureci, A. Imamoglu, Phys. Rev. A \textbf{81}, 061801(R) (2010)

\bibitem{Dario2} S. Ferretti, L.C. Andreani, H.E. Tureci, D. Gerace, Phys. Rev. A \textbf{82}, 013841 (2010).

\bibitem{Michal} Michal Grochol, Phys. Rev. B \textbf{79}, 205306 (2009).


\bibitem{Yamamoto} Neil Na, Shoko Utsumomiya, Lin Tian, and Yoshihisa Yamamoto, Phys. Rev. A \textbf{77}, 031803(R) (2008).

\bibitem{Dario3} S. Schmidt, D. Gerace, A. A. Houck, G. Blatter, H. E. Tureci, Phys. Rev. B \textbf{82}, 100507(R) (2010).

\bibitem{SimulIon} P. A. Ivanov, S. S. Ivanov, N. V. Vitanov, A. Mering, M. Fleischhauer, K. Singer, Phys. Rev. A \textbf{80}, 060301(R) (2009).


 
 \bibitem{Glass} M. P. A. Fisher, \textit{et.al}., Phys. Rev. B \textbf{40}, 546 (1989); Pinaki Sengupta, Stephan Haas, Phys. Rev. Lett. \textbf{99}, 050403 (2007); Ehud Altman, Yariv Kafri, Anatoli Polkovnikov, and Gil Refael, Phys. Rev. Lett. \textbf{100}, 170402 (2008).


\bibitem{DIfluid} Long Dang, Massimo Boninsegni, and Lode Pollet, Phys. Rev. B \textbf{79}, 214529 (2009).

\bibitem{SMFT2} Ulf Bissbort, Ronny Thomale and Walter Hofstetter, Phys. Rev. A \textbf{81}, 063643 (2010).
 
\bibitem{SMFT} U. Bissbort and W. Hofstetter, EPL, \textbf{86} 50007 (2009).

 
 
\bibitem{Pola} Michael Knap, Enrico Arrigoni, Wolfgang von der Linden, Phys. Rev. B \textbf{81}, 104303 (2010); Peter Pippan, Hans Gerd Evertz, Martin Hohenadler, Phys. Rev. A \textbf{80}, 033612 (2009).

\bibitem{Nega} A. Peres, Phys. Rev. Lett. \textbf{77}, 1413 (1996); G. Vidal and R. F. Werner, Phys. Rev. A \textbf{65}, 032314 (2002).


\bibitem{AtomEntCritical} M.X. Huo, Ying Li, Z. Song, C.P. Sun, Phys. Rev. A \textbf{77}, 022103 (2008).




\bibitem{PhyRealEns} H. M. Wiseman and John A. Vaccaro, Phys. Rev. Lett. \textbf{87}, 240402 (2001).

\bibitem{Nha} Hyunchul Nha and H. J. Carmichael, Phys. Rev. Lett. \textbf{93}, 120408 (2004).

\bibitem{Phys.RealEnt} Eduardo Mascarenhas, Daniel Cavalcanti, Vlatko Vedral and Marcelo Fran\c{c}a Santos, Phys. Rev. A 83, 022311 (2011); S. Vogelsberger and D. Spehner, Phys. Rev. A 82, 052327 (2010).

\bibitem{Carol}  M. C. O. Aguiar, V. Dobrosavljevi\'c, E. Abrahams e G. Kotliar, Phys. Rev. Lett. 102, 156402 (2009)

\bibitem{Fazio} D. Rossini and R. Fazio, Phys. Rev. Lett. \textbf{99}, 186401 (2007).








\end{thebibliography}
\end{document}